\documentclass{sigchi-ext}
\usepackage[T1]{fontenc}
\usepackage{textcomp}
\usepackage[scaled=.921]{helvet} 
\usepackage{graphicx} 
\usepackage{balance}  
\usepackage{booktabs} 
\usepackage{ccicons}  
\usepackage{ragged2e} 
\usepackage{xcolor}

\def\plaintitle{SIGCHI Extended Abstracts Sample File: Note Initial
  Caps} 
  
\def\emptyauthor{}
\def\plainkeywords{Human-AI Interaction; Mental Models; Artificial Intelligence}

\title{Understanding Mental Models of AI through Player-AI Interaction}

\numberofauthors{2}
\author{%
  \alignauthor{%
    \textbf{Jennifer Villareale}\\
    \affaddr{Drexel University} \\
    \affaddr{Philadelphia, PA} \\
    \email{jmv85@drexel.edu} }\alignauthor{%
    \textbf{Jichen Zhu}\\
    \affaddr{IT University of Copenhagen}\\
    \affaddr{Copenhagen, Denmark}\\
    \email{jichen.zhu@gmail.com} }\alignauthor{%
} }

\definecolor{linkColor}{RGB}{6,125,233}
\hypersetup{%
  pdftitle={\plaintitle},
  pdfauthor={\emptyauthor},
  pdfkeywords={\plainkeywords},
  bookmarksnumbered,
  pdfstartview={FitH},
  colorlinks,
  citecolor=black,
  filecolor=black,
  linkcolor=black,
  urlcolor=linkColor,
  breaklinks=true,
}

\begin{document}

\CopyrightYear{2021}
\setcopyright{rightsretained}
\conferenceinfo{HCXAI '21}{HCXAI '21: ACM CHI Workshop on Operationalizing Human-Centered Perspectives in Explainable AI, May 08--09, 2021, Virtual Event}
\isbn{978-1-4503-6819-3/20/04}
\doi{https://doi.org/10.1145/3334480.XXXXXXX}

\maketitle

\RaggedRight{}

\begin{abstract}
Designing human-centered AI-driven applications require deep understandings of how people develop mental models of AI. Currently, we have little knowledge of this process and limited tools to study it. This paper presents the position that AI-based games, particularly the player-AI interaction component, offer an ideal domain to study the process in which mental models evolve. We present a case study to illustrate the benefits of our approach for explainable AI.



\end{abstract}

\keywords{\plainkeywords}

\ccsdesc[500]{Human-centered computing~Human computer interaction (HCI)}
\ccsdesc[300]{Human-centered computing~Haptic devices}
\ccsdesc[100]{Human-centered computing~User studies}


\section{Introduction}
Artificial intelligence (AI), including traditional AI and machine learning (ML), has been used to provide a wide range of user experiences (UX) such as adaptive voice user interface~\cite{myers2019impact} and automatic medical diagnosis~\cite{cai2019human}. Researchers have argued that the complexity of AI-driven experiences makes them particularly challenging for both users~\cite{myers2018patterns,kliman2020adapting,torkamaan2019can} and designers~\cite{dove2017ux,zhu2018explainable}. These problems stem from the uncertainty around AI's capabilities and how well it may perform, as AI systems may demonstrate unpredictable behaviors that can be disruptive or confusing~\cite{yang2020re,amershi2019guidelines}. Designers face the difficult challenge of anticipating how users perceive and encounter the AI's capabilities, as this is fundamental to ensure a positive user experience~\cite{yang2020re}. 

Understanding the mental models of users has shown to improve the UX of digital applications~\cite{norman1983some,staggers1993mental,de2017they}. For AI-driven experiences, understanding users' mental models of the AI and designing the UX accordingly can be instrumental to developing human-centered approaches to AI explainability~\cite{gero2020mental} and human-AI interaction in general~\cite{rutjes2019considerations,amershi2019guidelines,kliman2020adapting,zhu2021player}. A mental model approach goes beyond users’ preferences and information needs associated with a given explanation and provides a rich picture of how users comprehend a system. Understanding this process can help address the open problem of \textit{how} users want to interact with XAI and \textit{when} users may need an explanation~\cite{abdul2018trends,ehsan2019automated}.

Current literature on mental models of AI is relatively limited. Most existing work studies users' mental models {\em after} interacting with an AI~\cite{bansal2019beyond,kulesza2012tell}. However, an essential part of interacting with AI-driven applications is to continuously modify mental models based on the response of the AI. Yet, we have little knowledge of how it happens and limited tools to study this process.



The central position of this paper is that {\em AI-based games, particularly the player-AI interaction component~\cite{zhu2021player}, offer an ideal domain to study the process in which mental models evolve.} Games have a long history of getting users to interact with AI in a variety of forms such as non-player characters~\cite{villareale2020innk}, procedural content generation~\cite{Valls-Vargas2017,shaker2016procedural}, experience management~\cite{sharma2010drama,zhu2019experience} and personalized adaptive games~\cite{ram2007artificial,Valls-Vargas2015}. Further, games have long been used to study how users form mental models~\cite{riemer2016impacts,graham2006acognitive,gero2020mental,furlough2018mental}, research has also shown that games provide a context that encourages the development of sophisticated mental models~\cite{furlough2018mental}. 
It is well established that games can offer a safe and motivating environment for experimentation and failure, which is important to the development of an accurate mental model. By contrast, the high stake domains that many AI-driven applications are used (e.g., loan approval, medical diagnosis) incentivize users to stick to known safe approaches and thus limit the deepening of their mental models and the observable behaviors for researchers to analyze. Games provide users the freedom to act while also providing researchers the rich data necessary to compare how, and if, a user's mental model has evolved and potentially trace which design elements caused this change.

Our approach adds to existing research methods. ``In-the-moment'' development of mental models is notoriously difficult to study~\cite{gero2020mental,langan2004mental,franco2000grasping}. We argue that the gameplay data itself offers an inroad to users' mental models. Commonly used research methods, such as interviews and surveys, can interrupt user experience when applied during the interaction. Think-aloud protocols have been used effectively in games~\cite{gero2020mental}, but it is limited because mental models are often unconscious and thus hard to capture through conscious responses alone. As we illustrate below, the traces from player-AI interaction can provide a complementary approach to capture and analyze how users' mental models develop. We show how this construct can offer rich insights into the different ways users comprehend a system and how this can advance the notion of human-centered XAI. 

\section{Related Work}
\subsection{Mental Models of AI}
Mental models are personal, internal representations of external reality that people use to interact with the world around them~\cite{craik1943nature,johnson1983mental,norman1983some}. HCI researchers have used this construct to understand how individuals think a system works or behaves~\cite{norman1983some}. Recently, a growing body of work in HCI has studied how users form mental models of AI systems. Bansal et. al~\cite{bansal2019beyond} look at the effect of different kinds of AI errors on people’s mental models, using performance as an indicator of the {\em accuracy} of a mental model. Kulesza et al.~\cite{kulesza2012tell} studied the effect of accurate mental models on usability and satisfaction in the context of a music recommender system. More related, Gero et al.~\cite{gero2020mental} recently studied mental models of AI in a cooperative word guessing game. They found that people tend to revise their mental models in the face of anomalies. Their work demonstrates the potential of using games to study mental models of AI. However, their findings are based on data collected in think-aloud and survey methods. This paper takes a step further and argues that the gameplay data itself offers an inroad to users' mental models, in addition to conscious responses.

\subsection{Failure in Games}

Failure is an important step in the process of acquiring accurate mental models. As argued above, one advantage of using games to study mental models is the ability to fail safely. In most cases, game designers design failure to improve players' knowledge of the game and to facilitate understanding of the problem at hand~\cite{gee2005learning,juul2013art,anderson2018failing}. For instance, Jesper Juul~\cite{juul2013art} discusses that failure provides players with the opportunity to consider the \textit{why} and reveal the depth behind the system they interact with. Gee~\cite{gee2005learning} describes good games as learning machines. He argues that failure is central to learning and is part of the fun. 

While we know that failure can be good for reflection and learning~\cite{juul2013art,gee2005learning}, we do not yet know how to design failure to help users develop more accurate mental models of AI. One possible approach is to utilize attribution theory~\cite{juul2009fear,kelley1967attribution}. Based on this theory, the three main targets a person may assign failure are to a person, entity, or circumstance. Juul ~\cite{juul2009fear} used it to analyze how failure re-adjusts player perceptions of a game. Building on Juul's work, we propose to use attribution theory as a first step to pinpoint where the mental model revision is taking place. 

\section{Understanding Mental Models of AI Through} 
\section{Gameplay}
Games are particularly well-suited to study the development of mental models of AI for a few reasons. First, it is well established that user behavior is more permissive in games when compared to real-life environments~\cite{frazier2012team,williams2010mapping}. However, this flexibility can shed light on the instinctive development of mental models as gamified environments have limited long-term consequences. This freedom can provide insights into the full breadth of possible interactions and misconceptions, and in turn, better tailor explanations.
Second, games are typically designed as a cyclic experience in which the core gameplay loop gets progressively more challenging. The built-in repetition and progression of complexity provide researchers the rich data necessary to evaluate a player's mental model and potentially trace which design elements caused this change as players are more likely to act on a hypothesis in gamified environments.  
Third, the play traces from the gameplay (e.g., behavioral data on how players interact with an AI-based character) can provide informative ``snapshots'' of their mental models at a given moment. For example, how an experienced {\em StarCraft} player competes with an AI opponent encodes her mental model of how the AI works and what its limitations include. Using games to study mental models of AI can help XAI consider how to best leverage AI explanations in ways that support the mental model process.



\subsection{Player-AI Interaction Framework}
\begin{marginfigure}
\hspace*{-0.2in}
  \begin{minipage}{\marginparwidth}
    \centering
    \includegraphics[width=0.9\marginparwidth]{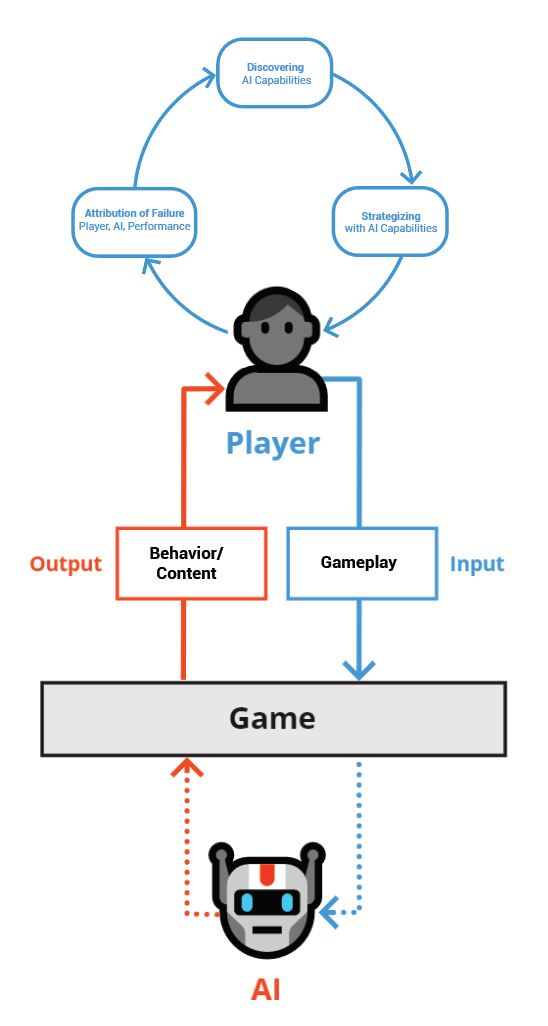}
    \caption{This framework displays the player-AI interaction and player's mental model process of 1) attributing failure, 2) discovering the AI's capabilities, and 3) strategizing with this knowledge in the game.}~\label{fig:cycle}
  \end{minipage}
\end{marginfigure}

Our framework (Figure~\ref{fig:cycle}) includes three key elements: the player-AI interaction, the player's attribution of failure, and the process for discovering and strategizing with the AI's capabilities. We adopt Swink’s feedback loops~\cite{swink2008game} and extend Aytemiz et al.'s~\cite{aytemiz2020diagnostic} failure taxonomy for games. Swink depicts interactivity between the human and the computer as a closed-loop where both actively listen, think, and respond. Aytemize et al.'s work extends this feedback loop to include where players fail during interaction with a game but does not consider how failure is attributed by the player. We adjusted their models by including the AI, specifically the player-AI interaction loop, and extend the player's mental model to include how they attribute failure (e.g., person (player), entity (AI), and circumstance (performance)), and strategize with the AI's capabilities in the game.

By analyzing the cycle of 1) player input (i.e., what the player does with the AI) and the AI output (i.e., the feedback the player receives) can provide a detailed snapshot into the player's mental model. This picture can provide insights into how users comprehend the AI and what response they may anticipate given their current understanding of the AI's capabilities. For instance, through interaction, players may discover a limitation of the AI and then use that knowledge to adjust their gameplay. By examining the changes in the players' gameplay we can gain insight into the process in which mental models evolve. In combination with how the player assigns failure, this can be used as an indicator of where the mental model revision is taking place and what the mental model includes in its revision.

\section{Case Study: \textit{Hey Robot}}
\textit{Hey Robot}~\cite{hey} is a multiplayer board game where three or more players take turns to get a smart speaker (e.g., {\em Amazon Echo}) to say a given word without saying the word themselves. For example, if a player chooses the word prompt ``{\em India},'' she may ask the smart speaker ``Alexa, where is the Taj Mahal?'' If the smart speaker's answer includes the word ``{\em India},'' the player wins the round. Over each round, players iteratively puzzle about what question is best to ask the smart speaker. To be successful in the game, the player must develop a basic understanding of the AI's capabilities. This game is particularly interesting for

our purpose because it forces players to develop their mental model of the AI through trial-and-error rather than pure reasoning. 

\textbf{Player-AI interaction.} In our preliminary study, we captured the player-AI interaction by collecting 1) the input (i.e., what questions players ask) and 2) the output (i.e., the response of the smart speaker) over a single game session. We used the player's question as an indicator of the player's ``in the moment'' mental model. The question illustrated what the player believes will work given the chosen word prompt. We then used the smart speaker's response to examine what factors may have influenced a change and what concepts the player may have discovered regarding the AI. This understanding can be observed in the strategies players employ, such as how they approach their questions for each round. 

Our initial analysis suggests that the development of players' mental models iteratively cycles through an exploration and elaboration phase. First, players would explore by structuring questions around related topics (Word: {\em Flying}, Question: "{\em Alexa, how does a bird travel?}") that made sense to them. Then, they cycled through a variety of related topics until one of them was successful. Through failure, players would pick up on a concept about the AI and this discovery would transition players to elaborate on this concept in the formulation of their next question. For example, players picked up that the AI could not understand sentences as a whole and might be latching onto keywords. Therefore, in the following rounds, they structured questions as if it were a web search by including keywords that would be recognizable to the AI (Word: {\em Jurassic Park}, Question: "{\em  Alexa, what \textbf{movie} was \textbf{Richard Attenborough} in during the \textbf{'90s}?}"). These moments encouraged the players to try and connect what they knew personally with the AI's output. This suggests that the player's mental model has been updated and that they are now strategizing with this new knowledge in the game. By using the player-AI interaction, we can now map the process of how and when mental models are revised and what factors may have influenced this change. This would otherwise not be possible to capture through alternative methods as players are unlikely to articulate all the subtle changes during gameplay.

\textbf{Players attribution of failure.} Between rounds, players would often engage in discussions with their teammates to explain why the smart speaker did not give the desired response. We analyzed the discussions to identify where the revision of the mental model focuses. For example, a player assigned failure to the AI specifically: "{\em the AI is just really bad at understanding me}". Other players would assign failure to themselves: "{\em I could have asked the question differently}" and would often further describe ways to improve their question for the following round.

Our initial analysis suggests that players who assigned the failure to themselves tended to pick up more concepts about the AI and elaborate on them. This can indicate that players who structure their mental model around what the AI can and can not do may revise their models more frequently. Other players that attributed failure specifically to the AI asked questions that made sense to them as opposed to forming a mental model with what the AI could respond to. This can indicate that these players need more scaffolding to refocus their mental model. By considering how players are assigning failure, we may better understand what the mental model includes in its revision.  

\textbf{Design implications for XAI.} Insights from AI-based games apply to XAI for a few reasons. First, games allow users to iterate on a variety of hypotheses and comfortably act on them. Capturing and analyzing how this process evolves freely can inform how to scaffold explanations and reveal the individual differences regarding how and when users may need an explanation. Second, games provide users a new way to interact with AI. To go beyond static explanations, HCXAI researchers can draw from games, as they provide a means to explore the system's behavior freely, to inform future work in interactive explanations~\cite{abdul2018trends}.

Based on our preliminary analysis, HCXAI researchers should further investigate how to help users shift their attention to the AI after a failed interaction. For instance, we observed players in the game formulate mental models without the AI's capabilities. Explanations could address this by supporting the redirection of attributions from the AI (i.e., entity) to the player (i.e., person) to effectively strategize with the AI's capabilities. Then, scaffolding explanations to gradually expose the users to other capabilities over time. 

Finally, researchers exploring interactive explanations should further investigate the concept of play when designing interactions. For instance, we observed this playfulness transform the smart speaker's inability to understand user command/intent into a source of fun and challenge for users to explore the boundaries of the AI's limitations. This perspective offers a useful design approach, complementary to the static, single message techniques~\cite{abdul2018trends}.

\section{Conclusion}
In this paper, we present our position that AI-based games, particularly the player-AI interaction, offer an ideal domain to study the process in which mental models of AI evolve. We identified the main benefits of this approach and illustrated them through a case study where we presented our preliminary findings. 

\balance{} 

\bibliographystyle{SIGCHI-Reference-Format}
\bibliography{references}

\end{document}